\documentstyle[11pt]{article}
\title{\begin{flushright}
{\normalsize
NSF-ITP-94-40\\
NUC-MINN-94/4-T \\
April 1994\\}
\end{flushright}
\bf BARYON - ANTIBARYON PRODUCTION BY DISORDERED CHIRAL CONDENSATES}
\author{ {\bf Joseph I. Kapusta} \\
  {\small \it School of Physics and Astronomy, University of Minnesota,
Minneapolis, MN 55455}\\
{\bf Ajit M. Srivastava} \\
{\small \it Institute for Theoretical Physics, University of California,
Santa Barbara, CA 93106}}

\date{}

\begin{document}

\maketitle

\begin{center}
Abstract\\
\end{center}

We investigate the production of baryons and antibaryons in the
central rapidity region of high energy nuclear collisions within
the framework of the Skyrme model
taking into account the effects of explicit chiral symmetry
breaking. We argue that the formation of disordered chiral
condensates may lead to enhanced baryon-antibaryon production
at low transverse momentum.

\vfill \eject

The possibility of producing quark-gluon plasma in
relativistic heavy ion collisions is very
exciting, especially from the point of view of observing the
chiral/confinement phase transition/crossover as the
plasma expands and cools. Recently, many investigations have
focussed on the possibility of the formation of domains in a
chiral phase transition where the chiral field may not be oriented
along the true vacuum.
Formation of a large domain with a {\em Disoriented Chiral Condensate}
(DCC) has been proposed by Anselm \cite{an}, by Blaizot and
Krzywicki \cite{BK} and by Bjorken, Kowalski and Taylor
\cite{bkt} in the context of high multiplicity hadronic collisions.
It was argued in \cite{bkt} and by Blaizot and Diakonov \cite{bd}
that, as the chiral field relaxes
to the true vacuum in such a domain, it may lead to coherent emission
of pions. A motivation for this proposal comes from
Centauro  events in cosmic ray collisions \cite{cntr}.
In the context of quark-gluon plasma, Rajagopal and Wilczek
proposed \cite{rw} that the nonequilibrium dynamics during the phase
transition may produce  DCC domains. They argued that long wavelength
pion modes may get amplified leading to emission of coherent pions.
Gavin, Gocksch and Pisarski have argued
\cite{ggp} that large domains of DCC can arise if the effective masses
of mesons are small, while Gavin and M\"uller propose \cite{lrg} the
annealing of smaller domains to give a large region of DCC.
Blaizot and Krzywicki \cite{fluct} have pointed out that even if
the average domain size is very small, random fluctuations could
result in some subset of all nuclear collisions having a large
domain of DCC.
Recently we have suggested that the effective potential of the linear
sigma model may have a second, local minimum at a chiral angle of $\pi$,
which could mimic some of the effects of a first order phase transition
even when the theory has no true thermodynamic phase transition of
any order \cite{us}.  We referred to this as a proximal chiral phase
transition.  We further argued that this could lead to large domains
of DCC.

In this paper we consider another signature which is expected from the
formation of initial domain structure.  This utilizes the Skyrmion
picture of the nucleon \cite{bal} and is based on the ideas
discussed in \cite{tom,ellis,ams}. In these previous
investigations, production of Skyrmions was studied in a
manner analogous to the production of cosmic strings and monopoles
in the early Universe. However, as we will elaborate later, the
investigations in \cite{tom,ellis} did not take into account
the boundary conditions which are needed for the specification of
Skyrmions. Also, the studies in \cite{tom,ellis,ams} corresponded
to the situation when explicit chiral symmetry breaking is absent.
As we will argue below, the formation of DCC
with the inclusion of chiral symmetry breaking terms
leads to enhanced production of baryons and antibaryons 
at low transverse momenta.

We use the sigma model in its linear representation.
The Lagrangian is expressed in terms of a scalar field $\sigma$ and
the pion field \mbox{\boldmath $\pi$}.
\begin{equation}
{\cal L} \,=\, \frac{1}{2}\left(\partial_{\mu}\sigma\right)^2
+ \frac{1}{2}\left(\partial_{\mu} \mbox{\boldmath $\pi$}\right)^2
-\frac{\lambda}{4} \left(\sigma^2 + \mbox{\boldmath $\pi$}^2 -
c^2/\lambda\right)^2 - V_{SB} \, .
\end{equation}
The piece of the Lagrangian which explicitly breaks chiral symmetry
is $V_{SB}$.  In the absence of this term, the potential has
the shape of the bottom of a wine bottle.  Chiral symmetry is
spontaneously broken in the vacuum, the pion is the massless Goldstone
boson, and the $\sigma$ meson gets a mass
on the order of 1-2 GeV. For the present purpose it is sufficient
to consider the symmetry breaking potential to have linear and
quadratic terms, as follows.
\begin{equation}
V_{SB} \,=\, - \sum_{n=1}^2 \frac{\epsilon_n}{n!}\sigma^n \,.
\end{equation}
(Even more general possibilities may be entertained \cite{us}.)
We have then at our disposal four parameters in the
effective Lagrangian: $\lambda, c, \epsilon_1$ and $\epsilon_2$.
These parameters must be restricted so as to give the proper pion
mass, pion decay constant, a reasonable value for the $\sigma$ mass,
PCAC in the weak field limit, and the condition that the ground
state of the theory occur at $\sigma = f_{\pi}$ and
$\mbox{\boldmath $\pi$} = {\bf 0}$.  We obtain
\begin{eqnarray}
m_{\pi}^2 &=& \lambda f_{\pi}^2 - c^2 \, ,\\
m_{\sigma}^2 &=& 2 \lambda f_{\pi}^2 + m_{\pi}^2 - \epsilon_2 \, ,\\
f_{\pi} m_{\pi}^2 &=& \epsilon_1 + \epsilon_2 f_{\pi} \, .
\end{eqnarray}
One may consider varying $\epsilon_1$ between 0 and $f_{\pi} m_{\pi}^2$.

To obtain the effective potential, we expand the fields about an
arbitrary point as
\begin{eqnarray}
\sigma(x) &=& v\,\cos\theta + \sigma'(x) \, ,\\
\mbox{\boldmath $\pi$}(x) &=& {\bf v}\,\sin\theta +
\mbox{\boldmath $\pi$}'(x) \, .
\end{eqnarray}
The primes denote fluctuations about the given point.  At one loop
order, and in the high temperature approximation, the effective
potential is \cite{us}
\begin{equation}
V(v, \theta; T) \,=\, \frac{\lambda}{4}v^4 - \frac{1}{2}
\left( c^2 + \epsilon_2 \cos^2\theta - \frac{\lambda T^2}{2}
\right) v^2 - \epsilon_1 v \cos\theta \, .
\end{equation}
Here $T$ is the temperature.
Although the aforementioned approximations
may not be quantitatively accurate, the above expression
at least provides a basis for discussion.
In particular, when $\epsilon_1 > 0$ the bottom of the
potential is tilted.  If $0 < \epsilon_1 < f_{\pi} \epsilon_2$ then
there is a second, local, minimum located at $\theta = \pi$ which
lies higher than the global minimum at $\theta = 0$.
In this case, a (unstable) domain wall would separate the
two minima. In this paper we will primarily consider the case
when there is only one minimum of the effective potential
and only briefly comment on the case when a second local
minimum is also present.

Now let us consider the Skyrmion.
Quite simply, it is a configuration where the chiral field
winds non-trivially around the manifold of degenerate minima
of the effective potential at zero or finite temperature
(neglecting for the time being the symmetry breaking terms),
which is $S^3$ for the two flavor chiral model.
An important aspect of the Skyrmion configuration is that it requires
the chiral field to approach a constant value at large distances.
This compactifies the local spatial region to a three-sphere $S^3$.
As the chiral field, when restricted to the minimum of the potential,
is valued in order-parameter space (which is $S^3$ for the two flavor
case as mentioned above) one can have nontrivial winding number
configurations on the spatially compactified $S^3$.

It was first suggested in \cite{tom}, and later developed in
\cite{ellis}, that a domain structure in
quark-gluon plasma may lead to the formation of Skyrmions.
However, the asymptotic boundary conditions which are required
to define an Skyrmion were not considered in \cite{tom,ellis}
which led to an over-estimate of baryon production (for the
case when there are no symmetry breaking terms present).
Theoretical investigations with proper treatment of boundary
conditions  \cite{ams,txt1,txt2} show that it is practically
impossible to form a full Skyrmion by domain formation which has
integer winding number right at the time of formation (in contrast
to the estimates in \cite{tom,ellis}). However, as discussed in
\cite{ams,txt1,txt2}, one can calculate the probability of forming
partial winding number Skyrmions which could later evolve into a
full Skyrmion. These studies showed that such $partial$ Skyrmions
form with a probability of about 0.05 in a suitable region. This
region was taken to be equal to four elementary domains in
\cite{ams}. The reason for taking four domains is due to the boundary
conditions and corresponds to the extended nature of the Skyrmion
\cite{ams}.  Numerical simulations give the probability of
the formation of (partial) textures, which are analogs of the
Skyrmions in the context of the early Universe, to be about
0.04 \cite{smln}. As we have mentioned above, all these previous
estimates corresponded to the case when there are no explicit
symmetry breaking  terms present in the effective potential.
We will discuss this case first and
then proceed to estimate Skyrmion production for the case when
the effective potential is tilted.

The essential idea behind these estimates is based on a method first
proposed by Kibble in the context of the production of topological
defects during phase transitions in the early Universe \cite{kbl}.
It has been applied to analogous phase transitions in condensed
matter physics \cite{nlc} and to the study of the
formation of glueballs and baryons/antibaryons in models based on QCD
strings \cite{carl}. In the context of quark-gluon plasma, it leads
to the following picture. As the phase transition proceeds, the chiral
field settles to the minimum of the potential. In a small region
the variation of the field will be small to avoid gradient energy.
However, in regions which are far from each other, the orientation
of the chiral field may differ. This situation can then be
represented by triangulating the region containing the plasma
in terms of elementary domains whose vertices are separated by a
distance $\xi$. Here $\xi$ represents some sort of correlation
length beyond which the orientation of the chiral field
varies randomly. For the chiral phase transition one could take this
distance to be about 1 fm. Vertices of the domains  then are associated
with the chiral field which is roughly uniform close to a given
vertex but has randomly varying orientation (but constant
magnitude, assuming the phase transition is completed) from one
vertex to another. In-between
any two adjacent vertices the variation of the chiral field will
be smooth to minimize the gradient energy. This last requirement is
generally called  the geodesic rule as this requires that the chiral
field traces the shortest path on the order parameter space $S^3$ 
between any two nearest vertices.

With these conditions we now estimate the probability
of Skyrmion formation. To get a clear picture, consider first
a lower dimensional example.  A Skyrmion in two space dimensions arises
when the  field takes values in $S^2$.  A Skyrmion configuration
in this case corresponds to the field winding completely on $S^2$
in some area in the two dimensional plane.  Divide the two dimensional
space into triangular domains. One such domain is
shown in figure 1. The triangle we want to consider has
its center $P_0$ a distance $\xi$ away from its vertices.
The values of the Skyrmion field at $P_1, P_2, P_3$ are $\alpha_1,
\alpha_2, \alpha_3$ which define a patch, shown as the shaded area on $S^2$
in the figure, when these points are joined by the shortest path on $S^2$
and the resulting enclosed area is filled. Consider the image of this patch
formed by diametrically opposed points of
each point in this patch. If the value of the field $\alpha_0$
at $P_0$ is such that it lies in this image
then the variation of the Skyrmion field in the triangular area will cover
more than half of $S^2$. For such a configuration it is natural to expect
that the dynamics will force the field to develop full winding on $S^2$,
resulting in a Skyrmion \cite{ams,txt1,txt2,smln}.  The probability
for the formation of such a configuration can be estimated to be 1/8
because the average area of the patch is
1/8 of the total area of the two-sphere. To see this we note that
the three points $\alpha_i$ trace three great circles on $S^2$ when
these points are joined pairwise by shortest geodesics. These three
great circles divide $S^2$ into eight spherical triangles leading
to the average area of one triangle equal to 1/8 of the surface area
of the sphere \cite{txt2}. We note that this is the same as the
probability of getting a monopole configuration in three space dimensions
corresponding to the winding of $S^2$ into $S^2$. Indeed, a Skyrmion
in $d$ space dimensions
is just a stereographic projection of a field configuration corresponding
to winding of $S^d$ onto $S^d$ in $d+1$ space dimensions.  [Roughly
similar results can be obtained by discretizing $S^2$ in term of four
points and requiring that the field at $P_0, P_1, P_2, P_3$
all correspond to different points on this discretized $S^2$.]
This probability of 1/8 is for a Skyrmion extended over the entire
big triangle, which consists of three smaller triangles.  Therefore
the probability per unit area is $\xi^{-2}/6{\sqrt 3}$.

It is important to note that figure 1 shows only a partial winding
case; it is the dynamics which is supposed to evolve this
configuration into a full integer winding number Skyrmion. Numerical
simulations carried out in the context of textures in cosmology
have shown that  it indeed happens \cite{smln}. As we will
show later, one of the effects of tilting the potential by adding
explicit symmetry breaking terms is to enforce such an evolution
of the partial winding into the full winding configuration.

Next, let us consider Skyrmions in three space dimensions. Figure 2 shows
the elementary domain: it is a tetrahedran whose vertices are a
distance $\xi$ away from its center $P_0$. Again, the values of
the Skyrmion field at the vertices $P_1, P_2, P_3, P_4$ of
the tetrahedran define a volume (${\cal V}$) on $S^3$
obtained by joining these values of the field
along the shortest path and filling the enclosed region.
The condition to get a partial winding which could evolve into
a full winding Skyrmion is that the field at $P_0$ be such
that it lies in the image of ${\cal V}$ obtained by considering
diametrically opposite points of ${\cal V}$. The average volume
of ${\cal V}$ is 1/16 of the total volume of $S^3$. Again, as in the
two dimensional case, this is  because the four minimal surfaces
formed by taking three $P_i$ at a time divide $S^3$ into 16 spherical
tetrahedra \cite{txt2}.  This gives
the probability of the Skyrmion formation to be 1/16.

We mention here that the issue of estimating Skyrmion formation
probability is not as straightforward as the estimation of
other topological defects, such as strings and monopoles.
For Skyrmions (and textures) dynamics plays a very important role.
This is clear when we note that we are calculating the probability for
only partial windings, and assuming that when this winding is more
than some critical value (say 0.5) then the field will evolve into an
integer winding Skyrmion. It could  very well be that many partial
winding Skyrmions may not be able to
develop into full winding Skyrmions. Further, our analysis
does not account for the field
configuration in neighboring domains which could affect
the evolution of the field in the domain under consideration.
There are several estimates of this probability in the literature
with resulting probabilities around 0.05 \cite{ams,txt1,txt2}.
Numerical simulations show this probability to be about
0.04 \cite{smln}. From these considerations, we will take the Skyrmion
formation probability to be about 0.04. This gives the probability
of Skyrmion formation per unit volume as
\begin{equation}
{\cal P} \,\simeq\, 0.08 \: \xi^{-3} \, .
\end{equation}
Taking $\xi$ to be 1 fm, the above probability gives the number of
Skyrmions as about 20 in a region of volume equal to 250 fm$^3$,
a typical quark-gluon plasma volume expected in heavy ion collisions.
As the probability of baryon and antibaryon formations
are equal, roughly half of these will be antibaryons. These antibaryons
should have very small transverse momenta; in fact they should
be practically at rest with respect to the local plasma. 

It is important to realize that these baryons and antibaryons will be 
produced $in~addition$ to those produced in thermal processes. The 
production of topological objects due to domain formation is a 
non-equilibrium
process and is independent of thermal production of such objects.
The only effect of thermally produced objects is to affect
the uniformity of neighboring domains. Therefore, thermal production
will only affect our estimates by a factor roughly equal to the
number of thermally produced objects per domain, which is
much smaller than one for baryons. Another
important aspect of this production mechanism is that it leads to
strong correlation between baryon and antibaryon formations.
This essentially arises from the fact that the formation of
a Skyrmion fixes the winding in the region common with
neighboring domains, enhancing the probability of formation of
opposite winding in those domains; see \cite{ams} for details.
Net baryon number of course should be fixed to be equal to the initial
baryon number of the plasma.  It is a non-trivial issue how to
incorporate exactly the constraint of net baryon number within the 
framework of this mechanism for Skyrmion production. Skyrmion
is a localized topological object and one can not just fix the
winding at the boundary of the region to control the net
Skyrmion number inside. In fact, it is very
interesting to consider baryon-rich matter, with a relatively
large net baryon number and associated chemical potential, and see
how this affects our estimates of the production of baryons and
antibaryons.  Here we have restricted our attention to the central
rapidity region which is expected to be nearly baryon-free.
We hope to address these topics in a future work.

There will, of course, be constraints coming from the net energy
available during the phase transition to form so many baryons.
However, as discussed in \cite{ellis}, this constraint is
not strong due to the strong temperature dependence of the
baryon mass  which decreases by about a factor of 0.45 as one
approaches the phase transition temperature. It was also argued
in \cite{ellis} that annihilation of baryon-antibaryon pairs
before coming out of the region may also not be very effective.
We will not discuss these issues any further. This is because,
as we have discussed above, the estimation of the probability
of formation of the Skyrmion itself is very nontrivial issue and
to make more quantitative predictions requires detailed numerical
simulation.

So far we have ignored explicit chiral symmetry breaking,
which favors $\theta = 0$.  We will
now argue that this {\em increases} the probability of baryon production.
For this analysis it is sufficient to consider only the parameter
$\epsilon_1$ non-zero. Our analysis will remain essentially the
same even if $\epsilon_2$ is non-zero as long as the effective
potential does not develop a second local minimum (or the slope
of the minimum of the potential does not vary strongly as one
goes along the valley of the potential). At the end of the paper, we
will briefly comment  on the case when there is a second minimum
also present. Consider again a small
tetrahedran like the one shown in figure 2 but with sides equal to $\xi$.
The Skyrmion field at its vertices can be taken to be randomly
varying, but we can no longer assume that at any point in the center of
this triangle the Skyrmion field will have independent orientation.
If the potential was not tilted, then any distribution of the Skyrmion
field on these vertices would have been trivial in the sense that
it will only cover a small region on the order parameter space $S^3$ 
and hence will shrink, decreasing the initial partial winding to zero. 
This would not result in the formation of a Skyrmion.

However, when the potential is tilted the situation is different.
Assume that the four vertices of the tetrahedron are such that,
when we use the geodesic rule to determine the field configuration
for the inside of the tetrahedran, one of the interior points
corresponds to $\theta = \pi$.  As $\theta = 0$ is the absolute minimum,
the chiral field at $P_1, P_2, P_3$ and $P_4$ will start evolving to
approach $\theta = 0$.  However, the interior of the tetrahedran
covers the $\theta = \pi$ region and hence can not evolve to the
true minimum. [This is assuming that the patch on the order parameter
space $S^3$ is not too small,
otherwise it may shrink while rolling down. This depends on the magnitude
of tilt of the potential. Such small patches will decrease the following
estimate of the probability slightly. However, we will neglect it as
we are interested in rough estimates. Clearly, if the tilt is very small
then most of the patches will shrink and in the limit of zero tilt
the results of \cite{ams,smln,txt1,txt2} will be recovered.]

 As the field at the four vertices approaches $\theta = 0$,
the Skyrmion field will be forced to wind around $S^3$ completely,
giving us a full integer winding number Skyrmion configuration.
This entire argument is essentially the same
as the one used earlier for the case of untilted potential
in the following sense. Previously, we
required that the chiral field in the interior of the tetrahedran
(at $P_0$) should lie in the image of the volume ${\cal V}$
formed by its diametrically opposed points. This led
to the covering of most of $S^3$ formed by ${\cal V}$ and the field
at $P_0$. Instead, we now require that  ${\cal V}$  be such that
a special fixed value ($\theta = \pi$) lies inside it rather
than in its image. This means
that a small patch on $S^3$ formed only by ${\cal V}$ is sufficient.
Clearly it does not change the calculation of the probability.
However, this probability now applies to a
Skyrmion which extends over a {\em smaller}
tetrahedran, leading to a {\em larger} probability per unit volume.
[For the two dimensional case this
means taking a small triangle with sides $\xi$ and requiring
that the shaded area on $S^2$, corresponding to this triangle,
cover the $\theta = \pi$ point.  The probability calculation again
is unchanged but one gets larger probability per unit area.]
Thus, for a tilted potential, we get the probability of Skyrmion
formation per unit volume in three space dimensions as
\begin{equation}
{\cal P^\prime} \,\simeq\, 0.33 \: \xi^{-3} \, .
\end{equation}
With $\xi =$ 1 fm this leads to about 82 baryons and antibaryons
in a region with volume equal to  250 fm$^3$. This probability is
about 4 times bigger as compared to the case when the
explicit chiral symmetry breaking terms are absent. [As we
discussed above this enhancement factor is applicable when the
tilt is not too small. The detailed dependence of the probability
on the magnitude of tilt is not easy to estimate
and one needs numerical simulations to address this issue.]
We again emphasize that these baryons and antibaryons should be
produced in addition to those which are thermally produced.
These estimates of probability density depend on the choice
of the elementary domains. It is possible that in certain
situations different types of domains may be more appropriate,
resulting in a different estimate of the probability density.
For example, in a true first order phase transition proceeding by
nucleation of bubbles, spherical domains may be  more relevant.
[In such a case, instead of calculating probability per unit volume,
one should use probability per domain and multiply by
number of domains to get the net number of baryons.]

We briefly comment on the case when $\epsilon_2$ is also non-zero.
When the local minimum at $\theta = \pi$ is also present then the
chiral field may role down to either of the two minima in different
domains. Walls separating different domains then correspond
to different interpolating curves between the two minima. It
is possible that at the junctions of such walls a patch may form
on $S^3$ near the $\theta = \pi$ region which could evolve to
a full Skyrmion configuration as the field rolls down to the true
minimum. The probability of Skyrmion formation will then be
proportional to the number of junctions of domain walls.
[In fact, the structure of the Skyrmion itself may have novel features
in this case when the effective potential has a local minimum as well.]
It will be interesting to carry out the simulation of
the formation of Skyrmions, such as in \cite{smln}, but with the inclusion
of the explicit symmetry breaking terms to check how the Skyrmion
formation probability gets modified.

\section*{Acknowledgements}

This work was begun during the program {\em Strong Interactions at
Finite Temperature} at the Institute for Theoretical Physics in the
fall of 1993.  It was supported by the U.S. Department of Energy under
grant number DOE/DE-FG02-87ER40328 and by the U.S. National Science
Foundation under grant number PHY89-04035. We thank Carl Rosenzweig
and Trevor Samols for useful comments.

\section*{Figure Captions}

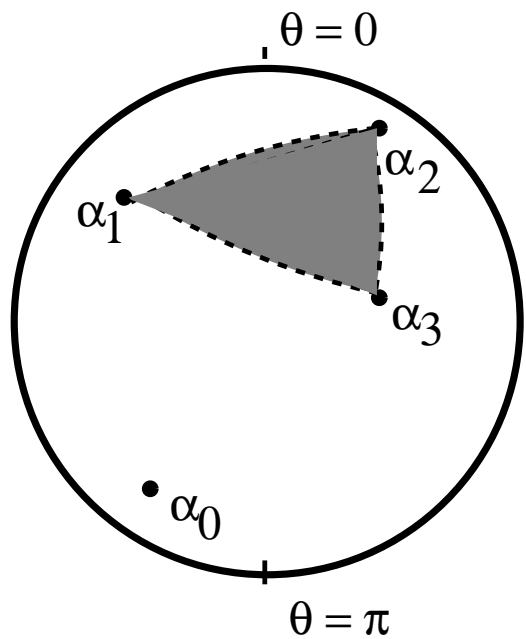
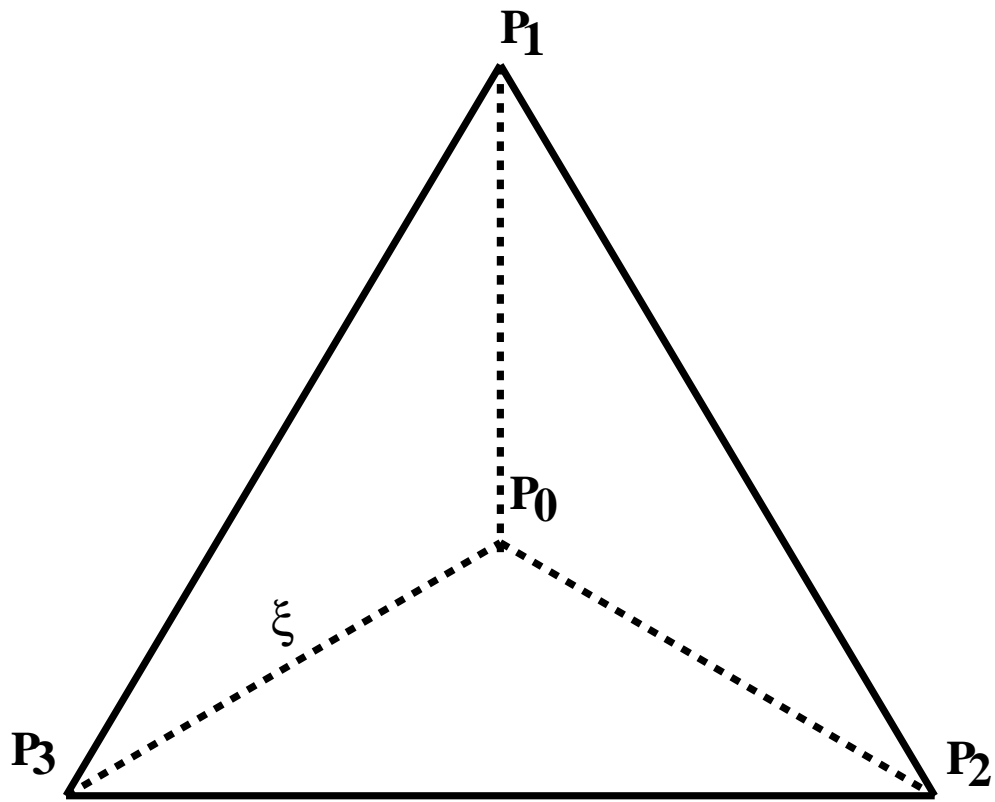
\noindent Figure 1: A tringular domain for the 2-dimensional case. $P_1, P_2,
P_3$ are a distance $\xi$ away from $P_0$. $\alpha_i$ represent the values
of the Skyrmion field on the manifold $S^2$. $\alpha_0$ should fall
in the diametrically opposite image of the triangular patch formed by
$\alpha_1, \alpha_2, \alpha_3$ for the field configuration on the triangular
domain to be able to evolve to a full Skyrmion.

\vspace{.25in}

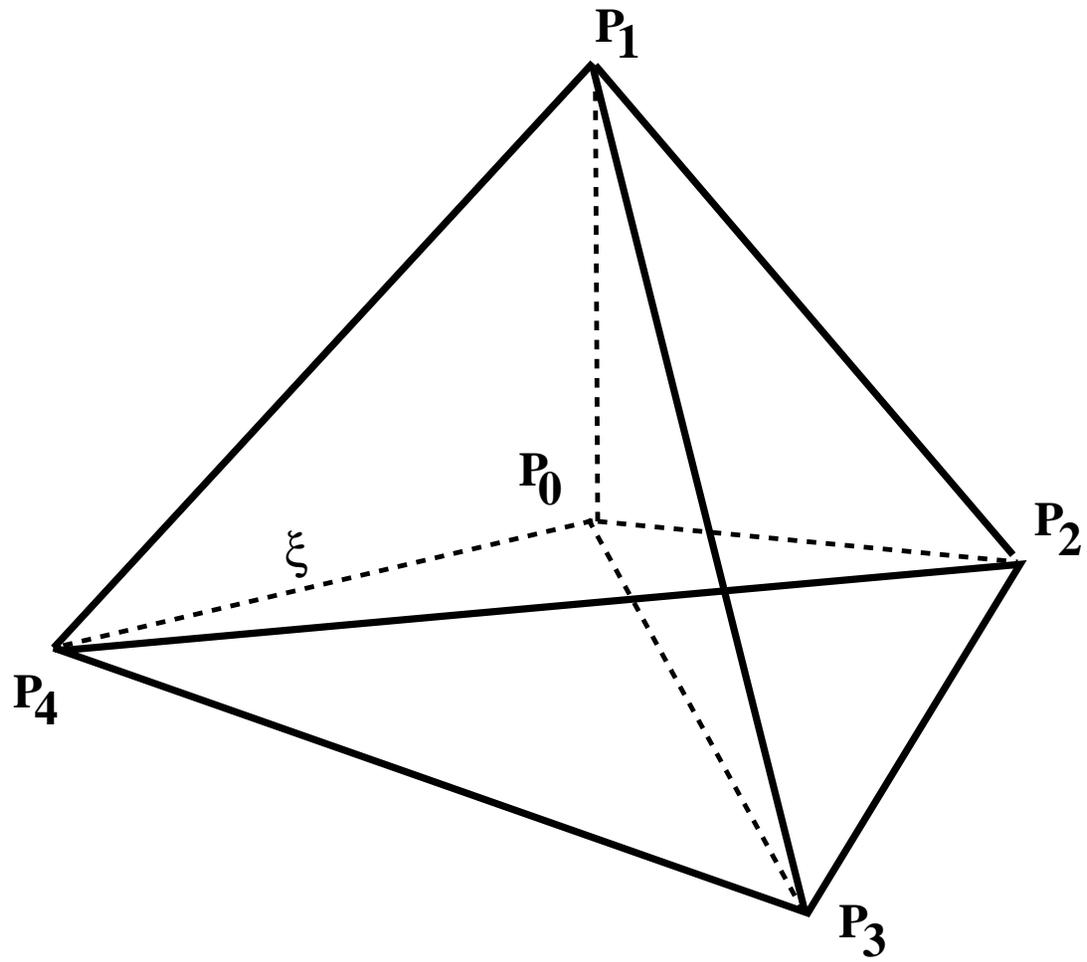
\noindent Figure 2: A tetrahderal domain appropriate for the Skyrmion
in 3-dimensions. All the vertices of the tetrahedran are a distance $\xi$
away from the point $P_0$ at its center.

\end{document}